# Records of sunspot and aurora during CE 960–1279 in the Chinese chronicle of the Sòng dynasty

Hisashi Hayakawa[1], Harufumi Tamazawa[2], Akito Davis Kawamura[2], Hiroaki Isobe[3,4]

[1] Graduate School of Letters, Kyoto University, Kyoto, Japan

[2] Kwasan Observatory, Kyoto University, Kyoto, Japan

[3] Graduate School of Advanced Integrated Studies in Human Survivability, Kyoto University, Kyoto, Japan

[4] Unit of Synergetic Studies for Space, Kyoto University, Kyoto, Japan

## Abstract

Records of sunspots and aurora observations in pre-telescopic historical documents can provide useful information about solar activity in the past. This is also true for extreme space weather events, as they may have been recorded as large sunspots observed by the naked eye or as low-latitude auroras. In this paper, we present the results of a comprehensive survey of records of sunspots and auroras in the *Sòngshǐ*, a Chinese formal chronicle spanning the tenth to the thirteenth century. This chronicle contains a record of continuous observations with well-formatted reports conducted as a policy of the government. A brief comparison of the frequency of observations of sunspots and auroras and the observations of radioisotopes as an indicator of the solar activity during corresponding periods is provided. This paper[†] is the first step of our project in which we survey and compile the records of sunspots and aurora in historical documents from various locations and languages, ultimately providing it to the science community as online data.

[†] This manuscript includes the original texts in Chinese, which is omitted in the version published in EPS.

## Background

Historical solar magnetic activity is of great interest because it provides crucial information on the origin of solar magnetic fields and its variability as well as the relationship of solar activity and the climate of the Earth (Usoskin 2013). Telescopic sunspot observations have been recorded since the era of Galileo about 400 years ago. The sunspot number is considered to be the standard index of the solar activity (Clette et al., 2014), and indeed, it is known as the longest-term scientific data (Owens 2013). For the older solar activity, cosmogenic radionuclides such as $^{14}C$ and $^{10}Be$ are commonly used as the proxy (e.g., Stuiver & Braziunas 1989; Beer et al. 1990). Eddy (1980) pointed out that pre-telescopic historical records of sunspots, auroras, and total eclipses (shapes of solar corona) could be useful for the reconstruction of the past solar activity. Several authors have published the lists of astronomical events including naked-eye sunspots and auroras in the literature from Japan (Kanda 1933; Matsushita 1956; Nakazawa et al. 2004; Shiokawa et al. 2005), Korea (Lee et al. 2004), Arabic countries (Basurah 2006), Europe (Fritz 1873; Link 1962; Dell'Dall'Olmo 1979; Stothers 1979; Vaquero & Trigo 2005; Vaquero et al. 2010), Russia (Vyssotsky 1949), North America (Broughton 2002), the Tropical Atlantic Ocean (Vázquez & Vaquero 2010), and China (Schove and Ho 1959; Keimatsu 1970–1976; Yau and Stephenson 1988; Saito & Ozawa 1992; Yau et al. 1995). A review on the historical records of solar activity is given in the monograph by Vaquero and Vázquez (2009).

In addition to the long-term changes, studies on historical solar activity have been attracting interest in the context of extremely intense solar flares and space weather events. Araki (2014) searched the historically largest geomagnetic sudden commencement (SC) using records of ground-based geomagnetic observations since 1868. The most intense geomagnetic storm on record is believed to be the so-called Carrington event in 1859 (Carrington 1859), whose estimated disturbance storm time (Dst) index is ~1760 nT (Tsurutani et al. 2003). The super-fast coronal mass ejection (CME) on July 23, 2012, is thought to have been as intense as the Carrington event, though it missed Earth (Baker et al. 2013, Russell et al. 2013). Another super-fast CME hit the Earth in August 1972 (Vaisberg and Zastenker 1976). This event yielded only moderate magnetic storm (Dst ~125 nT), presumably owing to the northward alignment of the interplanetary magnetic field, but it generated intense energetic particles and very rapid change in local geomagnetic fields (Tsurutani et al. 1992), causing significant damages in space and ground-based infrastructure (Lanzerotti 1992). If it were with a southward magnetic field, it could have produced very intense magnetic storm and aurora. Such intense energetic particles and magnetic storms severely damage space and ground infrastructure, resulting in a significant social and economic impact (NRC (National Research Council (U.S.) Committee on the Societal and Economic Impacts of Severe Space Weather Events) 2008, Royal Academy of Engineering Report 2013). Historical records of such extreme events may provide us some insight about what is the maximum strength of the flares and CMEs, and how frequent such extreme events are.

Recently, Maehara et al. (2012) discovered numerous super-flares in the solar-type stars (slowly rotating G-type stars), whose total energies are estimated to be $10^{33–35}$ erg, i.e., 10 to 1000 times more energetic than the most intense solar flares. Shibata et al. (2013) suggested that, theoretically, such super-flares may also occur on the present Sun. Interestingly, Miyake et al. (2012, 2013) found that the atmospheric $^{14}C$ content measured in tree rings exhibited marked increases during the Common Era‡ (CE) 774 to 775 and during CE 993 to 994, which indicate sharp increases in cosmic ray fluxes during those periods. The origin of these intense cosmic ray events is not known, but one potential cause is the solar energetic particles (SEPs) caused by extreme solar flares. Although the strongest SEPs in the modern observational records cannot account for these events, the SEPs from super-flares may be intense enough to cause these $^{14}C$ increments. Several studies have been published about the CE 774–775 including those by Usoskin et al. (2013), Cliver (2014), and Neuhäuser and Hambaryan (2014).

This paper is the first result of a project in which we aim to re-survey and compile the records of observations of sunspots and aurora from as many auroral sources as possible, and then provide the compilations as digital data to the scientific community. As a preliminary effort, in this paper, we present the data for the formal chronicle of the *Sòng* dynasty, *Sòngshǐ* (宋史, CE 960–1279), and we compare these results with data from cosmogenic radionuclide studies, such as the one performed by Miyake (2013).

# Methods

## Astronomical observations in the *Sòng* dynasty

In this study, we used the Chapter of Astronomy (天文志) in *Sòngshǐ*, mainly because it includes one of the richest sets of astronomical information among the chronicles of the Chinese dynasties. All the capitals in the *Sòng* era, where the observations were made, were located at latitudes between 30 N and 35 N. This period (CE 960–1279) overlaps with the CE 993–994 event as well as the so-called Medieval Warm Period (MWP: 10 CE–14 CE).

Among historical records worldwide, Chinese chronicles are remarkable for their feasibility as scientific data. Keimatsu (1970–1976) emphasizes the predominance of Chinese astronomical records because trained experts on astronomical observation made continuous observations at specified locations and took dated records of astronomical phenomena, often with detailed notes such as motions, shapes, and colors.

‡"Common Era" is a non-religious alternative of AD, namely "Anno Domini."

Thus, records in the Chinese official chronicles are regarded to be more objective than many other historical documents (Keimatsu 1970–1976), but we should keep in mind that Chinese astronomical observations were also made for the purpose of "fortunetelling" for policy makers. For example,

(a) 嘉泰四年二月庚辰夜，有赤雲間以白氣，東北亙天，後八日國有大火，占者以為火祥.

*Translation: on – March CE 1204 at night, red clouds (chìyún) appeared within white vapors, crossing the sky from the east to the west. After that, conflagrations occupied the country for eight days. Thus, astrologers regarded this as a symbol of fire*. (*Sòngshǐ*, Five Elements II b, p. 1413)

The reason for their consistent observations is based on Chinese culture and the politics of dynasties. Astronomical phenomena were traditionally thought to be signs from the heavens to the emperors reflecting their politics (i.e., stated at *Sòngshǐ*, Astronomical I, p. 949, translated text available in Appendix 1).

In order to deliver those heavenly messages to emperors, especially those of the *Sòng* dynasty, two observatories were constructed both in the imperial palace and near the capital city, *Kāifēng,* until the *Jìngkāng* Incident in CE 1126 (original text available in Appendices 2, 3, and 4, Yabuuchi 1967). Owing to their political importance, astronomical observations were made even during wars.

There is anecdotal evidence indicating the degree of sophistication and consistency of the astronomical observations in China during this period. During the invasion of the *Sòng* dynasty by the *Jīn* from the north, the Jīn captured *Kāifēng* in CE 1126. They broke all the observational instruments except for the spherical astrolabes, which were brought to *Běijīng* and adjusted by 4° in order to point to Polaris, as the instruments were moved from *Kāifēng* (35 N) to *Běijīng* (39 N) (the original text is available in Appendices 5–6).

After the fall of *Kāifēng*, the *Sòng* dynasty escaped southward via *Shāngqiū* (1128), *Yángzhōu* (CE 1128), and finally to *Línān*, the modern *Hángzhōu* (CE 1129–1279) (Mote 1999). It was in CE 1143 that the *Sòng* dynasty began reconstructing astrolabes. However, even during the period between CE 1126 and 1143 (without the use of astrolabes), we find records of both auroras and sunspots. This suggests that the *Sòng* dynasty continued observation even in such a chaotic time and without specially designed instruments (original text available in Appendix 6). Of course, we still have to note that frequency and accuracy of these observations must have been influenced by the political climate and other unknown factors.

## Search method

The *Sòngshǐ* includes descriptions of many kinds of phenomena observed in the sky. Since we are interested in the past solar activities, we searched for descriptions that could be regarded as records of sunspots or auroras. For this purpose, we used a search engine, *Scripta Sinica*, (http://hanchi.ihp.cinica.edu.tw) provided by *Academia Sinica* in Taiwan (http://www.sinica.edu.tw) that incorporates all the text data. This allowed us to automatically flag sentences that included keywords such as “black spot (黒子)” and “red vapor (赤氣),” which may refer to sunspots and red auroras, respectively. Once the sentences that include the keywords were picked up, we read corresponding parts of the original text to check whether they actually refer to sunspots or auroras. We also calculated the moon phase to determine the sky conditions for each date of observation.

## Sunspot records

Sunspots are described as black spots or black vapors in the sun or in terms such as "the sun was weak and without light" (Saito & Ozawa 1992). In the Astronomical Treatise of *Sòngshǐ*, these are categorized in the subsection of unusual phenomena in the sun. Since this period was long before the invention of telescopes, the observations were presumably made by the naked eye during sunrises/sunsets or through clouds.

Sunspot records are often accompanied with information about number, shape, and size. Examples are given as follows:

(a) 元豐元年十二月丙午，日中有黑子如李大，至丁巳散.

*Translation: on 11 January CE 1077, in the sun were black spots (hēizǐ) as large as plums. They disappeared on 22.* (*Sòngshǐ*, Astronomy, p. 1087)

(b) 紹興十五年丙午，日中有黑氣往來.

*Translation: on -- June CE 1145, in the sun were black vapors (hēiqì) shifting back and forth.* (*Sòngshǐ*, Astronomy, p. 1088)

The sizes are expressed by comparison with something tangible. In the above example, it was "as large as plums." The other expressions of size included are peach, glass, hen's egg, and duck's egg. Although we usually think of peaches as larger than plums, we do not know the relative and absolute size indicated by these expressions.

## Auroral records

Although the ancient Chinese did not know the physical nature of the phenomena, there are numbers of records that can be considered as the observations of auroras. For example, Keimatsu (1969a, 1969b) and Saito and Ozawa (1992) showed that the word "vapor" is likely to indicate auroras. Keimatsu (1970–1976) listed all the luminous phenomena seen at night and also discussed whether they correspond to auroras or not.

Similarly to Keimatsu, we assume that the records of luminous phenomena observed at night are potentially those of auroras. Therefore, we surveyed the words that refer to luminous phenomena, such as vapor, light, and cloud. In the Astronomical Treatise of *Sòngshǐ*, these are categorized in the subsection of clouds and vapors. From the list of the potential auroral candidates, we manually removed the following two types. The first are those without dates. Most of these are found in the chapters explaining how fortunetelling is performed, which includes conversations between the emperor and servants or sages. Therefore, they are unlikely to be the direct records of the observations. The second type includes those seen during the daytime. Some of the daytime phenomena are presumably the halo around the sun.

Records of auroral candidates often include information about their color, motion, and direction; length, shape, and number of their bands; and sometimes the location of the observation when it was not made in the capital city. Examples of auroral records are given as follows:

(a) 至道二年二月丙子夜，西方蒼白色氣長短八如彗掃．稍經天漢，參錯如交蛇.

*Translation: on 26 February CE 996 at night, in the west were eight long and short bands of pale-white vapors (cāngbáiqì) like comets. They gradually passed the Milky Way, entwined like patterned snakes.* (*Sòngshǐ*, Astronomy, p. 1308)

(b) 宣和元年七月戊午夜，赤雲起東北方，貫白氣三十餘道.

*Translation: 21 August CE 1119, red clouds (chìyún) appear in the northeast direction running through 30 ways of white vapors (báiqì).* (*Sòngshǐ*, Astronomy, p. 1314)

The color is described as white, red, blue, yellow, or their mixture. The colors may be associated with what traditional Chinese called the *Wǔxíng* or Five Elements, in which the world consisted of five elements, namely metal, fire, wood, soil, and water, which correspond with colors of white, red, green, yellow, and black, respectively. Their motion and direction are usually given by the eight points of the compass.

Sometimes, these descriptions include information about constellations, planets, or the moon accompanying the auroras. Their lengths are given in units of “chǐ”(尺) or “*zhàng*”(丈). In the *Sòng* Era, 1 *chǐ* was equal to 30.72 cm and 1 *zhàng* was 10 *chǐ* (Tonami et al. 2006). At this moment, we do not know how the lengths expressed using these units correspond to what was actually seen in the sky.

The shape and number of bands are described only in some of the recorded events. The shapes are expressed in a figurative way, such as “like serpents” or “like silk textiles.”

### Moon phase calculation

The moon phase is a trait of the sky that we can calculate accurately for historic dates. To do this, we referred to the 6000-year catalog of moon phases with Julian dates found at the NASA Eclipse website (http://eclipse.gsfc.nasa.gov/phase/phasecat.html). This catalog is based on an algorithm developed by Meeus (1998).

## Results and discussion

### Overall result

The lists of the candidates of sunspot and aurora are shown in Tables 1 and 2, respectively. In total, we found 38 sunspot candidates (black spot or black vapor in the Sun) and 193 auroral candidates (vapor/cloud/light during the night). The 193 auroral candidates include 75 white, 58 red, 28 blue white, and 32 others. The list is also available via our website: http://www.kwasan.kyoto-u.ac.jp/~palaeo/.

Figure 1 shows the annual number of records. We do not have enough records to see any signature of the 11-year solar cycle, but we can observe some long-term modulations. Overall, the period of CE 1100–1200 looks more active than the remainder. In particular, the sunspot records cluster during the periods of CE 1100–1145 and CE 1185–1205, when many auroral candidates are observed as well. There are also many auroral candidates during CE 1160–1175, but no sunspot candidate was found during this period.

## Color of aurora

The colors of the auroras are of interest because the colors have some information about the intensity of the auroral substorm and hence the solar activity. Usual auroras show green color from the forbidden line of atomic oxygen at 557.7 nm in the middle altitude (100–150 km), and red color of a forbidden line of atomic oxygen at 630.0 nm in the high altitude (>200 km). In very intense auroras, blue, violet, or pink colors in the low altitude (80–100 km) are sometimes seen that come from the molecular nitrogen lines in the blue and red parts of the spectrum. These lines are mostly excited by energetic electrons injected from the magnetosphere (Chamberlain 1961). Moreover, during large geomagnetic storms, so-called stable auroral red arcs (SAR arcs) become visible in the mid latitude (Rees and Roble 1975, Kozyra et al. 1997). All the auroral candidates presented in this paper are associated with colors. Among the 193 auroral candidates, 75 events are described as white, 58 as red, and 28 as blue white. The other 32 events include yellow, blue white, gold, etc. As mentioned above, the ancient Chinese use the five colors associated with the *Wŭxíng*, namely white, red, blue, yellow, and black. There is no green in *Wŭxíng*. Therefore, we assume that white colors in the Chinese records correspond to the green color of the aurora from the oxygen 557.7-nm line. Of course, there are always possibilities that other phenomena such as clouds and atmospheric optics are included. Keeping this in mind and recalling that the observations are made in the latitude of 30 N–35 N, the fact that there are more white auroras than the red auroras in the Chinese record seems puzzling. In such low-latitude region, usually only the red auroras are seen. This is either because the usual aurora in the higher latitude is seen from the lower latitude so that only the red color in the higher altitude (>200 km) is visible above the horizon, or because the aurora is the SAR arc, in which case the pure red (without green) color may be visible even near the zenith.

The white (green) auroras in the record indicate that the usual auroral oval expanded as low as 30 N–35 N where the capital of the *Sòng* era was located. Here, we should also note that the geomagnetic latitude of China in the *Sòng* era could be higher than that in the present (Butler 1992). There are some modern observational records that the auroral oval expanded to such a low latitude during the extremely intense geomagnetic storms (Loomis 1860, Kimball 1960). Therefore, it is likely that the green auroras were observed in the capital of *Sòng* when the solar activity was strong. However, that there are more red auroras than white (green) auroras still remains puzzling.

## Naked-eye sunspots

During the *Sòng* era, the sunspot observations are likely to have been made by the naked eye. Therefore, only very large sunspots could have been detected. Schaefer (1991, 1993) developed a theoretical model of sunspot visibility, including naked-eye observations. The simplest limit of visibility can be expressed as ~1.22 $\lambda/d$ ($\lambda$ and $d$ are the mean wavelength of sunlight and diameter of the optical system) (Vázquez & Vaquero 2010). Assuming a value of 1.5 mm for a diurnal observation, a value of 500 nm for the wavelength of the sunlight, and observers with standard eyesight (20-20 vision), the minimum angle of a naked-eye sunspot will be 70″, and a sunspot larger than that or as large as one thousandth the area of the solar hemisphere can be regarded as naked-eye sunspots.

As a previous study (Heath 1994) suggests, the number of such large sunspots is a good proxy of the overall solar activity. Figure 2 shows the number of large (greater than one thousandth of the area of the hemisphere) sunspots, as well as the sunspot number (the Wolf number or the international sunspot number) in the most recent three-and-a-half solar cycles. This figure demonstrates that the number of naked-eye sunspots is in fact a good proxy for the sunspot number.

Indeed, the emergence of large (naked-eye) sunspots is likely to be the necessary condition for the occurrence of extremely intense solar flares. Shibata et al. (2013) presented a scaling law that relates the energy of solar (and stellar) flares and the size of sunspots; for example, the maximum energy of a solar flare of a naked-eye sunspot can be around $10^{35}$ erg ($10^3$ times larger than our scientific observation record).

## Comparison with $^{14}C$ data

Using the $^{14}C$ content in tree rings, Miyake (2013) reported that there is a sharp increase reported in cosmic ray flux during CE 993–994. We did not find any candidates of sunspot nor aurora in these years. However, as can be seen in Fig. 1, we found a cluster of auroral candidates several years after this event. The closest aurora recorded is in 996 and there is a record of a sunspot in CE 1005. In this chronicle, we could not detect records clearly associated with the CE 994 event of rapid $^{14}C$ increase to which Miyake (2013) referred. Figure 3 shows $^{14}C$ data from Miyake (2013) in dots with error bars and counts of *Sòngshǐ* auroral records in bars. According to Fig. 3, in CE 994, neither aurora nor sunspot observation was recorded, but after the $^{14}C$ peak of CE 994, the number of observed auroras increased. There are two interpretations for this. The first is that this CE 994 event is not related to a short-time solar activity such as a flare. The second is that the record that involves the CE 994 event is not recorded in the *Sòngshǐ* for various reasons: weather, lack of observation, lack of importance for Chinese astronomers, etc., but the increase in the solar activity during that period is reflected in the increased auroral records some years later. This kind of difficulty is one of the limitations that our surveys frequently face.

We also note that auroras around CE 993–994 were observed in other parts of the world. In Ireland, the annals of Ulster reported a red-colored aurora observed in CE 992 (U992.4 of CELT 2000). In Korea, there is a record of a red aurora in CE 992 (Lee et al. 2004). In Germany and Denmark, three auroras in CE 992 and two auroras in CE 993 were reported (Fritz 1873). These may correspond to the astronomical event in CE 993–994.

## Auroras and moon phase

For studies of auroral records from historical articles, concerns on moon phase might be important in estimating the actual events indicated by the records. One reason for our interest in moon phase is that there are some natural optical events possibly expressed as vapor, such as "moon halo" and paraselene (or "moon dog"), which occur rarely with a nearly full moon and ice crystals in cold air. This kind of atmospheric optics would occur more frequently in a polar region, but may possibly occur at lower latitudes given appropriate weather conditions. If the frequency of the "potentially aurora" records depends on the moon phase, it indicates that there may be significant number of events that are actually atmospheric optics. In order to examine this, we calculate the moon phase (0 for the previous new moon and 1 for the next new moon) and make a histogram of historical records of vapors from *Sòngshǐ*.

Figure 4 shows a histogram of the vapor records of *Sòngshǐ* where no significant tendency is found. This result may be different from the previous studies (Vaquero et al. 2003, Vaquero & Trigo 2005) of auroral observations in Europe, which mentions a marked tendency of increased auroral observation towards the last quarter of the moon phase. These previous studies conclude the possibility of this tendency due to moon illumination during the observation window from dawn to the midnight. On the other hand, the Chinese observations were made throughout the night, according to ancient Chinese articles. Therefore, no skewness of Chinese auroral records around the new moon supports the presumption of all-night observations.

## Conclusions

We have surveyed sunspot and aurora-like descriptions in the official chronicle of the *Sòng* dynasty during CE 960–1279. We found 38 sunspots and 193 auroral candidates during that period, with information of their size, color, etc. The lists are provided in this paper. This is the first step for the project in which we will survey the records of sunspot and aurora in a variety of historical documents and provide the data online for use by the scientific community. We welcome and encourage the use of our data as well as the contributions that provide more data from diverse historical sources.

## Authors' contributions

This research was performed with the cooperation of authors as given: HH made philological and historical contributions. HT and ADK made contributions on the interpretation and analysis of the database. HI supervised this study. All authors read and approved the final manuscript.

## Acknowledgements

We thank Dr. R Kataoka and anonymous referees for interesting comments and suggestions. This work was supported by the Kyoto University's Supporting Program for Interaction-based Initiative Team Studies "Integrated study on human in space" (PI: H. Isobe), the Interdisiplinary Research Idea contest 2014 by Center of Promotion Interdisiplinary Education and Research, and also by the "UCHUGAKU" project of the Unit of Synergetic Studies for Space, Kyoto University.

# Appendices

## Appendix 1: *Sòngshǐ*, Astronomy I, p. 949

夫不言而信, 天之道也, 天於人君有告戒之道焉, 示之以象而已. 故自上古以來, 天文有世掌之官, 唐虞羲, 和, 夏昆吾, 商巫咸, 周史佚, 甘德, 石申之流. 居是官者, 專察天象之常變, 而述天心告戒之意, 進言於其君, 以致交脩之儆焉.

*Translation: needless to say, the heaven is a tao. The heaven shows taos of warnings only with symbols. Therefore, from ancient times, each era had their specialists of astronomy such as …(list of personal names)… These specialists kept observing whether the heavens were normal or abnormal. Hence, they state it as a warning of the heavens and advise it to the emperor in order to fulfill the duty of a secretary to teach his emperor not to make mistakes.*

## Appendix 2: *Shenkuo*, Mengxibitan, VIII, p. 4 (Shen 1966)

國朝置天文院於禁中, 設漏刻, 觀天臺, 銅渾儀, 皆如司天監, 與司天監互檢察. 每夜天文院具有無謫見, 雲物, 禎祥, 及當夜星次, 須令於皇城門未發前到禁中, 門發後, 司天占狀方到, 以兩司奏狀對勘, 以防虛偽.

*Translation: the government put the Bureau of Astronomy (tiānwényuà) in the Imperial Court, setting clepsydras, observatories, and astrolabes equally in the Bureau of Astro-manager (sītiānjiān) and made them check each other with the Bureau of Astro-manager. Every night the Bureau of Astronomy recorded if they found unusual phenomena, clouds, or omens, and star charts on that night. They had to bring these records to the Imperial Court before the gates were opened. After the gates opened and reports were taken from the Bureau of Astro-manager,*

*both of these reports were compared with each other and, in this way, they tried to avoid false reports.*

## Appendix 3: *Sòngshǐ*, Bureaus IV, p. 3879

太史局: 掌測驗天文, 考定曆法. 凡日月, 星辰, 風雲, 氣候, 祥眚之事, 日具所占以聞…其別局有天文院, 測驗渾儀刻漏所, 掌渾儀臺晝夜測驗辰象.

*Translation: the Bureau of Divination (tàishǐjú) is responsible for astronomical observation and designing calendars. This bureau is famous for making divinations of everything about the sun, the moon, stars, winds, clouds, weathers, and omens… On the other hand, the Bureau of Astronomy examines astrolabes and clepsydras. This bureau is responsible for observatory recordings of astronomical omens every night.*

## Appendix 4: *Sòngshǐ*, Bureaus V, p. 3923

元豐(1078-85)官制行, 罷司天監, 立太史局, 隸祕書省.

*Translation: during the Yuánfēng years (CE 1078–85), the system of bureaus got changed. The Bureau of Astro-manager was abolished and the Bureau of Divination was established under the Ministry of Secretary.*

## Appendix 5: *Jīn shi*, Calendars II, p. 523

金既取汴, 皆輦致于燕, 天輪, 赤道牙距, 撲輪, 懸象, 鐘鼓, 司辰, 報刻, 天池水壺等器久皆棄毀, 惟銅渾儀置之太史局候臺. 但自汴至燕相去一千餘里, 地勢高下不同, 望筒中取極星稍差, 移下四度纔得窺之.

*Translation: Jīn (dynasty) had already taken Biàn (Kāifēng) and sent everything to Yàn (Běijīng). They broke all the astronomical instruments such as …(list of names of observational*

*instruments)… except for the astrolabe. They put this astrolabe in the Bureau of Divination of the observatory. However, as it is more than thousand li from Biàn to Yàn and their altitudes are not the same, they tried to find the polar star with this instrument and succeeded by putting it 4 degree down.*

### Appendix 6: *Sòngshǐ*, Astronomy I, p. 950

靖康之變，測驗之器盡歸金人．高宗南渡，至紹興十三年，始因秘書丞嚴抑之請，命太史局重創渾儀.

*Translation: in the Jìngkāng Incident (CE 1126), all astronomical instruments were plundered by the Jīn people. Emperor Gāozōng escaped southward and finally in CE 1143, with a strong request of the Ministry of Secretary, he ordered the Bureau of Divination to make an astrolabe again.*

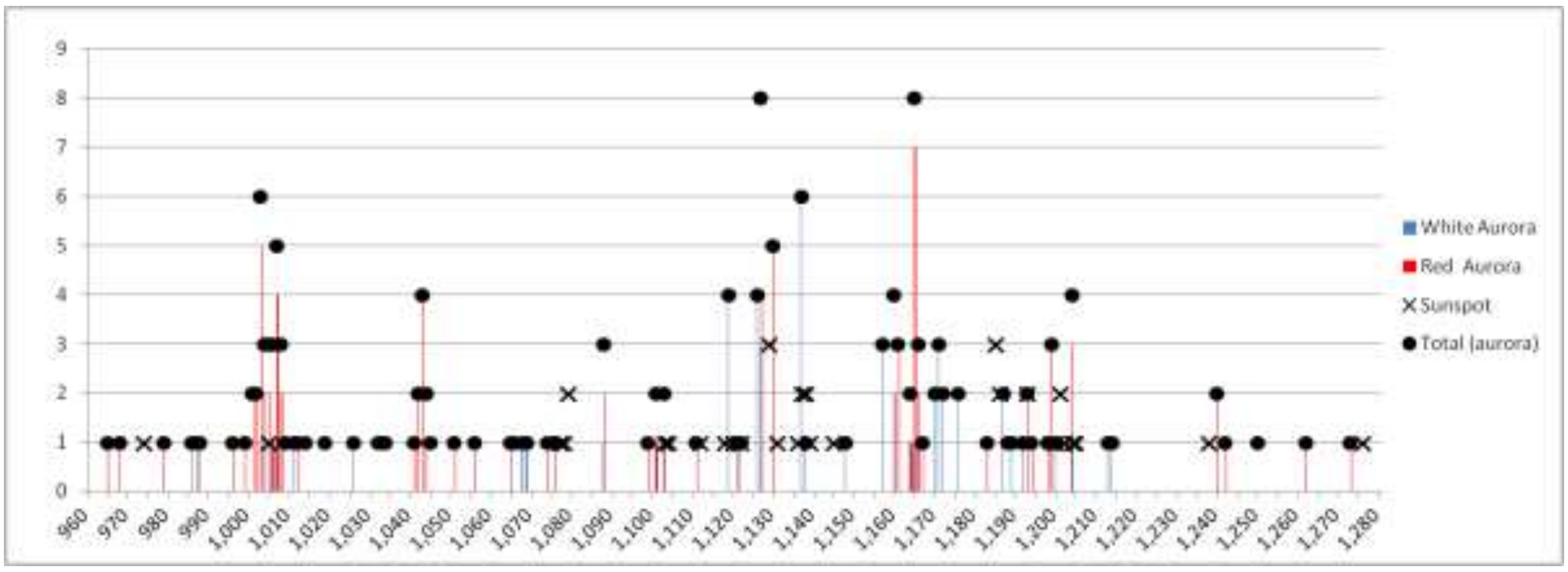

**Figure 1**: Change of the number of white vapor, red vapor, and black spots during the Sòng dynasty: The *bars in blue and red* represent the number of candidates of white and red aurora, respectively. The *black dots* are for the total number of aurora candidates. The *crosses* show the number of sunspots

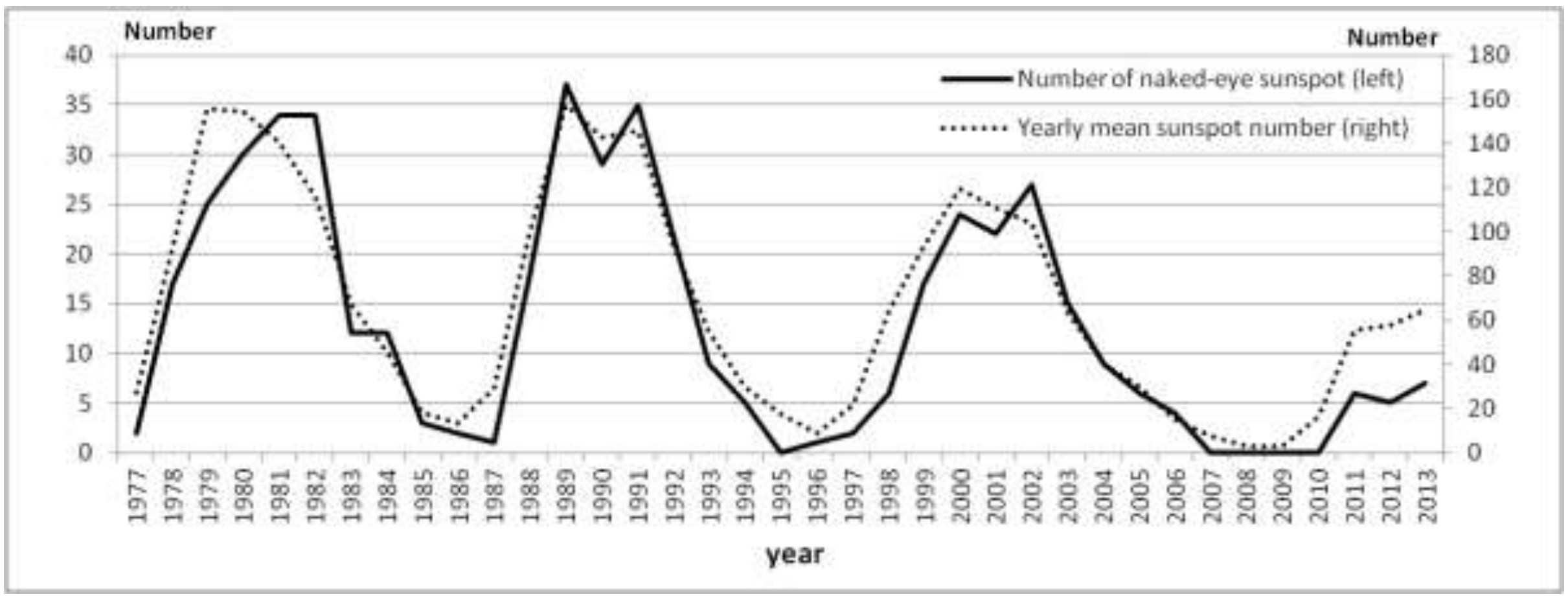

**Figure 2**: Change of the number over years for all and naked-eye sunspots: The number of naked-eye sunspots and the sunspot numbers are plotted with *solid line* and *dashed line,* respectively. The sunspot region data used for naked-eye sunspot calculation are retrieved from the NASA database (http://solarscience.msfc.nasa.gov/greenwch.shtml), which are originally given by the Royal Greenwich Observatory (RGO), US Air Force (USAF), and the US National Oceanic and Atmospheric Administration (NOAA). The yearly mean sunspot number is retrieved from WDC-SILSO, Royal Observatory of Belgium, Brussels (http://www.sidc.be/silso/datafiles)

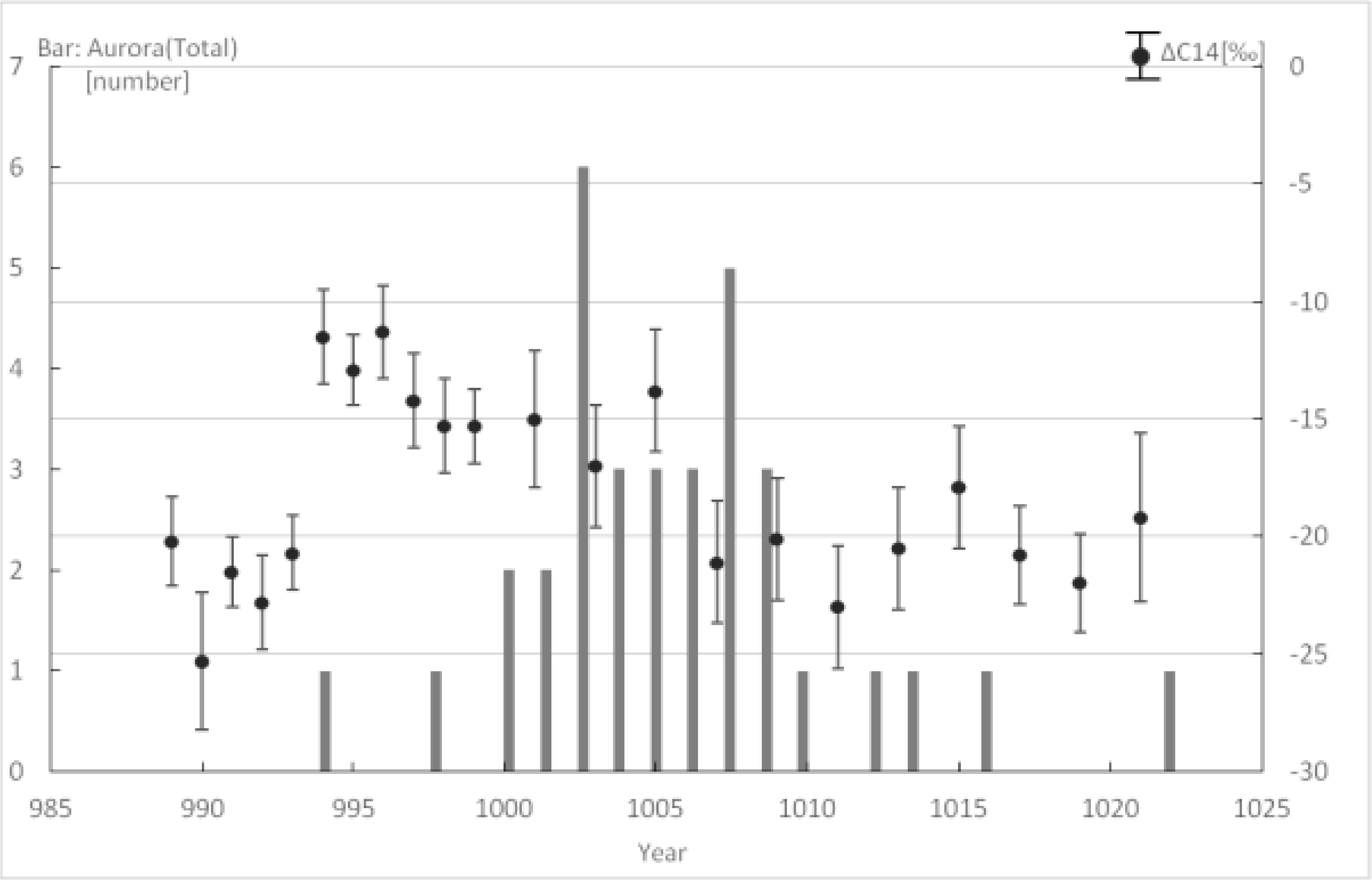

**Figure 3**: Comparison the change of $^{14}C$ in tree ring and the number of auroras. Miyake (2012) pointed out the spiky peak of $^{14}C$ at CE 774 (*right*). In Chinese chronicle, after CE 774, the number of observations of auroras increased. Between CE 985 and 1025, only in CE 1005 was the observation of sunspots recorded

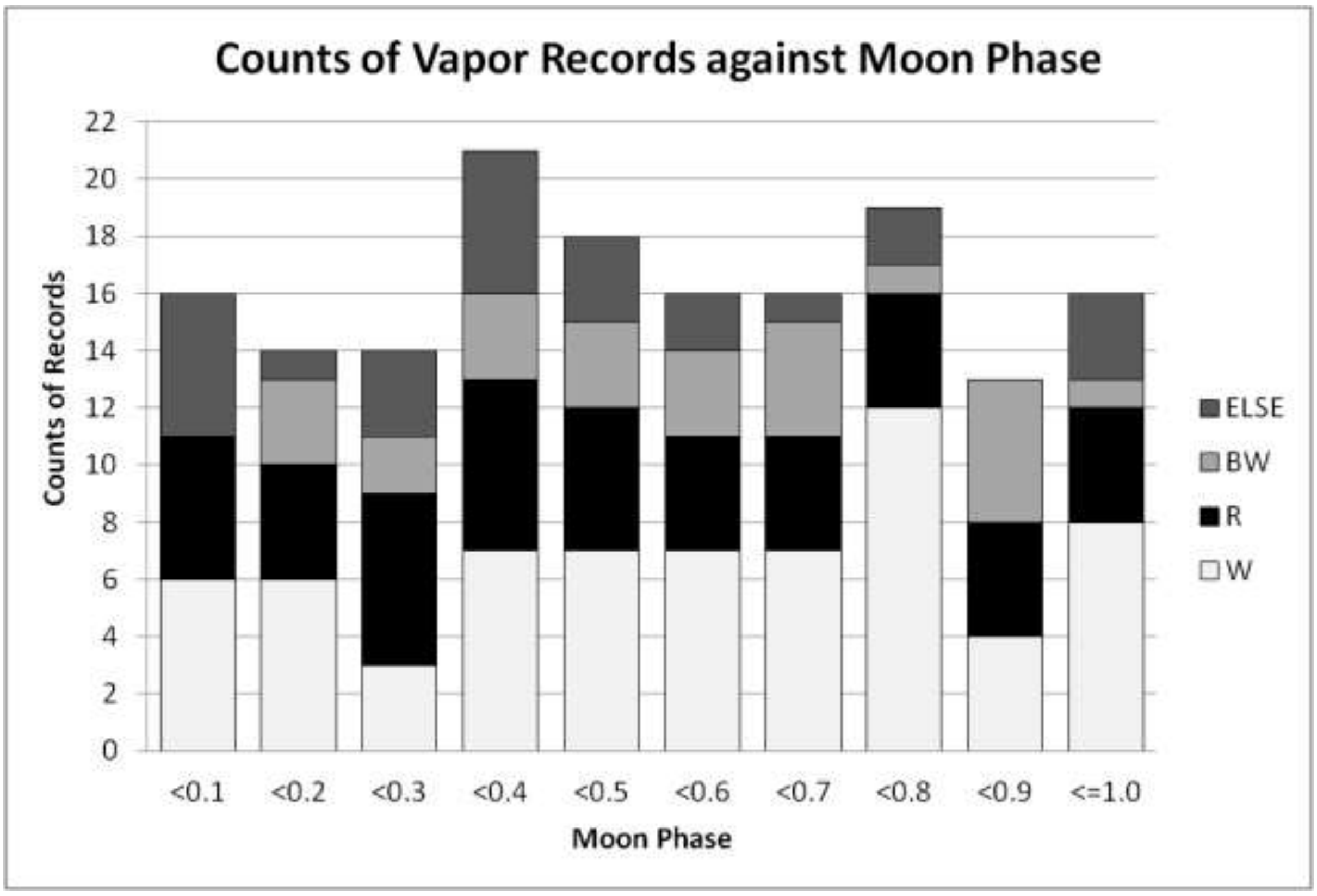

**Figure 4**: Counts of auroral candidates from *Sòngshǐ* against moon luminosity: This histogram is labeled with the reported color of vapor. The alphabetic labels represent the color of vapor as *W* (white), *R* (red), and *BW* (blue white)

**Table 1** Sunspot records

| Year | Month | Date | Description | Size | Counts | Place | Note |
|---|---|---|---|---|---|---|---|
| 974 | 3 | | BS | | 2 | Kāifēng | |
| 1005 | 2 | 6 | shade | | 2 | Kāifēng | Crescent-like shape |
| 1077 | 3 | 7 | BV | | | Kāifēng | Until 21 |
| 1078 | 3 | 11 | BS | | | Kāifēng | Until 29 |
| 1079 | 1 | 11 | BS | Plum | | Kāifēng | 12 days |
| 1079 | 3 | 20 | BS | Plum | | Kāifēng | |
| 1103 | 7 | 1 | WL | | | Kāifēng | |
| 1104 | 11 | | BS | Jujube | | Kāifēng | |
| 1105 | 11 | 6 | BS | | | Kāifēng | |
| 1112 | 5 | 2 | BS | Chestnut | | Kāifēng | |
| 1118 | 12 | 17 | BS | Plum | | Kāifēng | |
| 1120 | 6 | 7 | BS | Jujube | | Kāifēng | |
| 1122 | 1 | 10 | BS | Plum | | Kāifēng | |
| 1129 | 3 | 22 | BS | | | Línān | Until 4.14 |
| 1131 | 3 | 12 | BS | Plum | | Línān | 4 days |
| 1136 | 11 | 23 | BS | Plum | | Línān | Until 27 |
| 1137 | 3 | 11 | BS | Plum | | Línān | |
| 1137 | 5 | 8 | BS | | | Línān | Until 22 |
| 1138 | 3 | 16 | BS | | | Línān | |
| 1138 | 3 | 17 | BS | | | Línān | |
| 1138 | 11 | 26 | BS | | | Línān | |
| 1139 | 3 | | BS | | | Línān | For a month |
| 1139 | 11 | 20 | BS | | | Línān | |
| 1145 | 7 | | BV | | | Línān | |
| 1145 | 7 | | BS | | | Línān | |
| 1185 | 2 | 10 | BS | Jujube | | Línān | |
| 1185 | 2 | 15 | BS | | | Línān | Until 27 |
| 1185 | 2 | 27 | BS | | | Línān | |
| 1186 | 5 | 23 | BS | Jujube | | Línān | |
| 1186 | 5 | 26 | BS | | | Línān | |
| 1193 | 12 | 3 | BS | | | Línān | Until 12 |
| 1200 | 9 | 21 | BS | Jujube | | Línān | Until 26 |
| 1201 | 1 | 9 | BS | | | Línān | Until 29 |
| 1202 | 12 | 19 | BS | Jujube | | Línān | Until 31 |
| 1204 | 2 | 21 | BS | Jujube | | Línān | |
| 1205 | 5 | 4 | BS | | | Línān | |
| 1238 | 12 | 5 | BS | | | Línān | |
| 1276 | 2 | 17 | BS | Goose's egg | | Línān | |

Location of the observatories: *Kāifēng* (34.80 N, 114.30 E), *Lin'an* (30.25 N, 120.17 E)

*BS* black spots (黒子), *BV* black vapor (黒氣), *WL* the sun was weak and without light (日淡無光), *shade* (景)

**Table 2** Auroral records

| Year | Month | Date | Color | Description | Direction | Length | Counts | Place | Notes | Moon phase |
|---|---|---|---|---|---|---|---|---|---|---|
| 965 | 8 | 10 | W | V | w | 50 chǐ | | Kāifēng | | 0.36 |
| 968 | 11 | 2 | W | V | wn | 20 chǐ | 3 | Kāifēng | | 0.31 |
| 979 | 5 | 9 | W | V | wn | | | Kāifēng | | 0.38 |
| 979 | 5 | 19 | W | V | wn | | | Kāifēng | | 0.71 |
| 986 | 2 | 22 | R | V | n | | | Kāifēng | Until morning | 0.34 |
| 986 | 2 | | R | V | | | | Kāifēng | | |
| 987 | 2 | 10 | W | V | | | | Kāifēng | | 0.30 |
| 988 | 12 | | R | V | wn | Height 2 chǐ | | Kāifēng | | |
| 996 | 2 | 26 | BW | V | w | | 8 | Kāifēng | | 0.17 |
| 1001 | 4 | 20 | W | V | All | | 2 | Kāifēng | | 0.81 |
| 1001 | 8 | 20 | W | V | | | | Kāifēng | | 0.97 |
| 1003 | 5 | 13 | W | V | e-w | | | Kāifēng | | 0.33 |
| 1003 | 6 | 24 | W | V | | | | Kāifēng | | 0.73 |
| 1003 | 7 | 14 | R | V | | | | Kāifēng | | 0.42 |
| 1003 | 7 | 19 | W | V | | | Some | Kāifēng | | 0.59 |
| 1003 | 8 | 15 | W | V | ws | | | Kāifēng | | 0.50 |
| 1004 | 4 | | W-BW | V | | | 10~ | Kāifēng | | |
| 1004 | 6 | | W-BW | V | | | 10~ | Kāifēng | | |
| 1004 | 6 | 12 | W | V | | 7 chǐ | Some | Kāifēng | | 0.74 |
| 1004 | 8 | 17 | Y | V | | 5 zhàng | | Kāifēng | | 0.97 |
| 1005 | 2 | 28 | YW | V | | | | Kāifēng | Near the moon | 0.54 |
| 1005 | 3 | 21 | W | V | n | | 5 | Kāifēng | | 0.30 |
| 1005 | 11 | 5 | W | V | e-w | | | Kāifēng | | 0.04 |
| 1006 | 4 | 14 | R | V | n | | | Kāifēng | | 0.46 |
| 1006 | 4 | 14 | W | V | | | | Kāifēng | Near the moon | 0.46 |
| 1006 | 5 | | Y | V | | | | Kāifēng | Near the moon | |
| 1007 | 4 | 12 | W | V | e-w | | | Kāifēng | | 0.75 |
| 1007 | 4 | 13 | W | V | s | 2 zhàng | | Kāifēng | | 0.79 |

| | | | | | | | | | |
|---|---|---|---|---|---|---|---|---|---|
| 1007 | 4 | 23 | W | V | n | 10 zhàng | | Kāifēng | | 0.13 |
| 1007 | 5 | 13 | W | V | | 3 zhàng | | Kāifēng | | 0.82 |
| 1007 | 12 | 18 | R | V | m | 7 chǐ | | Kāifēng | | 0.23 |
| 1008 | 2 | 10 | Y | V | | | | Kāifēng | | 0.03 |
| 1008 | 2 | 24 | W | V | e-w | | 2 | Kāifēng | | 0.48 |
| 1008 | 8 | | W | CV | wn | | 30~ | Kāifēng | | |
| 1009 | 9 | 27 | Y | V | es | | | Kāifēng | | 0.21 |
| 1011 | 1 | 25 | BR | V | | | | Kāifēng | | 0.64 |
| 1012 | 2 | 28 | W | V | n | 5 zhàng | | Kāifēng | | 0.13 |
| 1014 | 6 | | | V | | | | Kāifēng | | |
| 1019 | 5 | 8 | Y | V | | | | Kāifēng | | 0.04 |
| 1026 | 10 | | R | V | w | | | Kāifēng | Sunset | |
| 1028 | 5 | 16 | BW | C | n-ws | Some | | Kāifēng | | 0.66 |
| 1032 | 11 | 7 | YW | V | | | 5 | Kāifēng | | 0.06 |
| 1033 | 1 | 28 | BW | V | wn | | | Kāifēng | | 0.85 |
| 1034 | 9 | 24 | BYW | V | | 7 zhàng | | Kāifēng | | 0.31 |
| 1041 | 9 | 1 | W | V | e | 10 zhàng | | Kāifēng | | 0.10 |
| 1041 | 9 | 24 | BW | C | wn | | | Kāifēng | | 0.89 |
| 1042 | 8 | 31 | W | V-C | n | | | Kāifēng | | 0.46 |
| 1042 | 12 | | BW | C | s | | | Kāifēng | | |
| 1043 | 3 | 13 | W | V | m-ws | 20 zhàng | | Kāifēng | | 1.00 |
| 1043 | 5 | 17 | W | V | wn-es | | 2 | Kāifēng | | 0.21 |
| 1043 | 9 | 23 | W | V | n | | | Kāifēng | | 0.58 |
| 1043 | 10 | 22 | W | V | m | 2 zhàng | | Kāifēng | | 0.58 |
| 1044 | 6 | | BW | V | | | 10~ | Kāifēng | | |
| 1044 | 10 | 16 | | V | en | 2 zhàng | | Kāifēng | | 0.77 |
| 1045 | 4 | 3 | Y | V | e | | | Kāifēng | | 0.45 |
| 1048 | | | W | V | n | | | Kāifēng | | |
| 1052 | 12 | 13 | W | V | n | 5 zhàng | | Kāifēng | | 0.66 |

| | | | | | | | | | |
|---|---|---|---|---|---|---|---|---|---|
| 1065 | 5 | 24 | W | V | wn-es | | | Kāifēng | | 0.57 |
| 1065 | 5 | 25 | W | V | w | | | Kāifēng | | 0.60 |
| 1066 | 7 | 19 | BW | C | e | 1 zhàng | | Kāifēng | | 0.82 |
| 1066 | 11 | | Y | V | | | | Kāifēng | | |
| 1067 | 3 | 2 | BW | C | s | 3 zhàng | 2 | Kāifēng | | 0.53 |
| 1067 | 4 | | BW | C | ws | 3 zhàng | 2 | Kāifēng | | |
| 1067 | 7 | 30 | W | C | en | 5 zhàng | | Kāifēng | | 0.57 |
| 1068 | 2 | 17 | BW | C | ws | 4 zhàng | | Kāifēng | | 0.38 |
| 1068 | 12 | | R | V | wn | | | Kāifēng | | |
| 1069 | 5 | 1 | BW | C | es | 3 zhàng | | Kāifēng | | 0.28 |
| 1069 | 8 | 9 | BW | C | en | 1 zhàng | | Kāifēng | | 0.66 |
| 1069 | 12 | | R | V | wn | | | Kāifēng | Every night this month | |
| 1072 | 7 | 27 | W | C | s | | | Kāifēng | | 0.33 |
| 1074 | 4 | 13 | BW-W | C-V | ws | 2 zhàng | | Kāifēng | Sunset | 0.48 |
| 1074 | 5 | 3 | BW | C | n | 52 chǐ | | Kāifēng | | 0.18 |
| 1074 | 5 | 13 | BW | C | wn | 3 zhàng | | Kāifēng | | 0.51 |
| 1074 | 7 | 4 | BW | C | s | 3 zhàng | | Kāifēng | | 0.27 |
| 1074 | 7 | 16 | BW | C | s | 2 zhàng | | Kāifēng | | 0.67 |
| 1074 | 7 | 17 | BW | C | e-s | 2 zhàng | | Kāifēng | | 0.70 |
| 1074 | 7 | 22 | BW | C | ws | 2 zhàng | | Kāifēng | | 0.87 |
| 1074 | 8 | 9 | BW | C | e | | | Kāifēng | | 0.47 |
| 1075 | 7 | 20 | BW | C | en | 4 zhàng | | Kāifēng | | 0.17 |
| 1076 | 7 | 10 | W | V | es | | | Kāifēng | | 0.23 |
| 1077 | 7 | 19 | BW | C | en | 3 zhàng | | Kāifēng | | 0.91 |
| 1079 | 5 | 13 | W | C | s | 3 zhàng | | Kāifēng | | 0.34 |
| 1079 | 5 | 16 | BW | C | s | 3 zhàng | | Kāifēng | | 0.45 |
| 1082 | 5 | 21 | BW | C | n | 2 zhàng | | Kāifēng | | 0.38 |
| 1088 | 7 | 24 | R-W | V | en | | | Kāifēng | | 0.14 |
| 1088 | 8 | 12 | R-W | V | en | | | Kāifēng | | 0.78 |

| | | | | | | | | | |
|---|---|---|---|---|---|---|---|---|---|
| 1088 | 8 | 13 | W | V | en | | | Kāifēng | | 0.81 |
| 1088 | 9 | 23 | R-W | V | n | | Some | Kāifēng | | 0.20 |
| 1099 | 10 | 15 | R-W | V | n | 5 chǐ | 10 | Kāifēng | | 0.97 |
| 1101 | 1 | 31 | R-W-Bk | V | en-w | | 4~ | Kāifēng | | 0.02 |
| 1103 | 1 | 8 | W | V | w | | | Kāifēng | Sunset | 0.94 |
| 1103 | 6 | 16 | BW | V | es | 3 zhàng | | Kāifēng | | 0.34 |
| 1111 | 12 | 17 | BW | V | | | | Kāifēng | | 0.55 |
| 1115 | 4 | 26 | W-Ra | C | n | | | Kāifēng | | 0.02 |
| 1117 | 6 | 29 | R-W | C-V | en | | | Kāifēng | | 0.92 |
| 1119 | 5 | 11 | R | V | | | 10~ | Kāifēng | | 0.01 |
| 1119 | 6 | 10 | R | V | wn | | | Kāifēng | | 0.03 |
| 1119 | 7 | 15 | R | V | n | | | Kāifēng | | 0.21 |
| 1119 | 8 | 21 | R-W | C-V | n | | 30~ | Kāifēng | Covered half sky | 0.47 |
| 1120 | 3 | 28 | R | C | en-wn | | | Kāifēng | | 0.79 |
| 1121 | 11 | 2 | BW | V | | 3 zhàng | | Kāifēng | | 0.82 |
| 1122 | 10 | 23 | R | V | w | | | Kāifēng | Sunset | 0.73 |
| 1125 | 5 | 15 | R | C | | | | Kāifēng | | 0.37 |
| 1126 | 2 | 4 | RW | V | w | | | Kāifēng | | 0.34 |
| 1126 | 10 | 3 | R | V | e | | | Kāifēng | Sunrise | 0.53 |
| 1126 | 11 | 19 | R | V | w | | | Kāifēng | Sunset | 0.11 |
| 1126 | 12 | 21 | R | V | | | | Kāifēng | | 0.22 |
| 1127 | 1 | 11 | W | V | | | | Kāifēng | | 0.90 |
| 1127 | 1 | | W | V | | | | Kāifēng | | |
| 1127 | 2 | 21 | R | L | wn | | | Kāifēng | | 0.27 |
| 1127 | 4 | 5 | W | V | ws-en | | | Kāifēng | | 0.74 |
| 1127 | 5 | | W | V | n | | | Kāifēng | | |
| 1127 | 9 | 20 | R | V | en | | | Kāifēng | | 0.42 |
| 1127 | 9 | 22 | R | V | en | | | Kāifēng | | 0.49 |
| 1127 | 9 | 27 | R | V | en | | | Kāifēng | At dusk | 0.65 |

| | | | | | | | | | |
|---|---|---|---|---|---|---|---|---|---|
| 1130 | 6 | | W | V | n | 10~ | Línān | | |
| 1130 | 6 | | R | L | en-es | | Dòngtíng | | |
| 1130 | 6 | | R-W | V-C | wn | 10~ | Línān | | |
| 1130 | 6 | 18 | R-W | C-V | n-es | | Línān | | 0.37 |
| 1130 | 7 | 5 | W | V | es | | Línān | | 0.95 |
| 1137 | 1 | 31 | R | V | en | | Línān | | 0.29 |
| 1137 | 2 | 14 | R | V | n | | Línān | | 0.76 |
| 1137 | 2 | 20 | R | V | en | | Línān | | 0.96 |
| 1137 | 3 | 4 | R | V | en | | Línān | | 0.42 |
| 1137 | 10 | | R | V | | | Línān | | |
| 1137 | 12 | 29 | R | V-C | s-en | | Línān | From the evening | 0.56 |
| 1138 | 10 | 6 | R | V | | | Línān | | 0.03 |
| 1148 | 8 | 17 | R | V | wn | | Línān | | 0.06 |
| 1157 | 3 | | R | V | | | Línān | | |
| 1157 | 4 | 30 | R | V | | | Línān | | 0.66 |
| 1157 | 8 | 15 | R | V | w | | Línān | Sunset | 0.30 |
| 1157 | 11 | 13 | R | V | w | | Línān | Sunset | 0.32 |
| 1160 | 2 | | R | V | en | | Línān | | |
| 1160 | 4 | 1 | R | V | en | | Línān | | 0.77 |
| 1160 | 12 | 18 | W | V | | | Línān | | 0.65 |
| 1160 | 12 | 19 | W | V | ws | | Línān | | 0.68 |
| 1161 | 1 | 2 | W | V | | | Línān | | 0.08 |
| 1161 | 12 | 21 | W | V | e-w | | Línān | | 0.12 |
| 1161 | 12 | 22 | W | V | | | Línān | | 0.15 |
| 1164 | 1 | 21 | W | V | ws | | Línān | | 0.86 |
| 1164 | 2 | 22 | W | V | | | Línān | | 0.95 |
| 1164 | 11 | 24 | R | C | w | | Línān | Sunset | 0.28 |
| 1165 | 3 | 4 | W | V | wn | | Línān | | 0.71 |
| 1165 | 5 | 1 | W | V | e-w | | Línān | | 0.66 |
| 1165 | 5 | 11 | W | V | | | Línān | | 0.99 |
| 1165 | 5 | 30 | BW | V | wn-en | | Línān | | 0.65 |
| 1165 | 6 | 3 | W | V | n | | Línān | | 0.77 |
| 1165 | 6 | 7 | W | V | | | Línān | | 0.91 |
| 1165 | 9 | 12 | R | V | m | | Línān | From sunset to evening | 0.18 |
| 1165 | 11 | 18 | BW | CV | s | | Línān | | 0.43 |
| 1165 | 12 | | R | C | w | | Línān | | |
| 1165 | 12 | 25 | W | V | e-w | | Línān | | 0.75 |
| 1166 | 1 | | W | V | e-w | | Línān | | |
| 1166 | 12 | 24 | W | V | | | Línān | | 0.02 |
| 1167 | 1 | | W | V | | | Línān | | |
| 1170 | 12 | 3 | R | V | e | | Línān | Sunrise | 0.81 |
| 1170 | 12 | 10 | R | V | w | | Línān | Sunset | 0.05 |
| 1171 | 9 | 1 | R | V | w | | Línān | Sunset | 1.00 |
| 1171 | 11 | 3 | R | V | w | | Línān | Sunset | 0.13 |
| 1171 | 11 | 4 | R | V | w | | Línān | Sunset | 0.16 |
| 1171 | 11 | 17 | R | V | w | | Línān | Sunset | 0.61 |
| 1172 | 10 | 28 | R | V | w | | Línān | Sunset | 0.33 |
| 1172 | 10 | 29 | R | V | e | | Línān | Sunrise | 0.36 |
| 1176 | 9 | 29 | R | V | w | | Línān | Sunset | 0.82 |
| 1176 | 9 | 30 | R | V | e | | Línān | Sunrise | 0.85 |
| 1183 | 2 | 16 | W | V | ws | Width 6 zhàng | Línān | From northern horizon to southern horizon | 0.75 |
| 1187 | 12 | 17 | R | V | w | | Línān | Sunset | 0.54 |
| 1187 | 12 | 18 | R | V | e | | Línān | Sunrise | 0.57 |
| 1188 | 9 | 29 | RY | V | s | | Línān | | 0.24 |
| 1189 | 1 | 15 | R | V | en | | Línān | Sunrise | 0.90 |
| 1192 | 2 | | R | V | | | Línān | | |
| 1193 | 12 | 5 | R-W | V-C | | | Línān | | 0.35 |
| 1193 | 12 | 6 | R-W | V-C | | | Línān | | 0.38 |
| 1194 | 7 | 2 | W | V | | | Línān | | 0.44 |
| 1194 | 7 | 9 | W | V | | | Línān | | 0.68 |
| 1198 | 9 | 17 | W | V | | | Línān | | 0.49 |
| 1199 | 3 | 9 | W | V | en | | Línān | | 0.42 |
| 1199 | 8 | 26 | W | V | en | | Línān | | 0.11 |
| 1199 | 9 | 7 | W | V | | | Línān | | 0.51 |
| 1200 | 11 | | R | V | | | Línān | | |
| 1204 | 3 | 29 | R | V | | | Línān | | 0.90 |
| 1204 | 3 | | R-W | C-V | en | | Línān | 8 days | |
| 1204 | 12 | 5 | W | V | | Some | Línān | | 0.43 |
| 1204 | 12 | 16 | W | V | | | Línān | | 0.79 |
| 1213 | 12 | 2 | R | V | e | | Línān | Sunrise | 0.63 |
| 1214 | 1 | 7 | R | V | w | | Línān | Sunset | 0.82 |
| 1240 | 3 | 16 | W | V | | | Línān | | 0.73 |
| 1240 | 5 | 13 | W | V | | | Línān | | 0.69 |
| 1242 | 3 | 3 | W | V | | | Línān | | 0.02 |
| 1242 | 5 | 3 | W | V | | | Línān | | 0.07 |
| 1250 | 12 | 10 | Rb | V | | | Línān | | 0.53 |
| 1262 | 8 | 16 | W | V | | | Línān | | 0.01 |
| 1273 | | | W | V | | | Xiāngyáng | | |

Length: *chǐ* (尺, 1 *chǐ* = 30.72 cm), *zhàng*(丈, 1 *zhàng* = 10 *chǐ*) (Tonami et al. 2006). Location of the observatories: *Kāifēng* (34.80 N, 114.30 E), *Línān* (30.25 N, 120.17 E), *Dòngtíng* (29.32 N, 112.95 E), *Xiāngyáng* (32.02 N, 112.16 E)

Color of auroral candidates: *W* white, *R* red, *B* blue, *Y* yellow, *Bk* black, *R-W* red and white, *RW* red white, *BW* blue white, *Ra* rainbow; types of recorded phenomena: *V* vapor (氣), *C* cloud (雲), *L* light (光); direction: *e* east, *w* west, *s* south, *n* north, *m* middle, *en* northeast; also, those abbreviations can be combined, for example *wn-es* from northwest to southeast.